\documentclass[pra,amsmath,aps,twocolumn,superscriptaddress]{revtex4}
\usepackage{amsmath,mathrsfs,amsbsy,amssymb,lmodern,graphicx,bm,amsthm,amsfonts}
\usepackage{units}
\usepackage{bbm}
\usepackage{multirow,color}

\newcommand{\idol}{\ensuremath{\mathbbm 1}}

\begin{document}
\title{Hierarchy of multipartite nonlocality in the nonsignaling scenario}
\author{Xiaoxu Wang}
\affiliation{College of Physics, Optoelectronics and Energy, Soochow University, Suzhou, 215006, China}
\author{Chengjie Zhang}
\email{zhangchengjie@suda.edu.cn}
\affiliation{College of Physics, Optoelectronics and Energy, Soochow University, Suzhou, 215006, China}
\affiliation{Key Laboratory of Quantum Information, University of Science and Technology of China, CAS, Hefei, 230026, China}
\author{Qing Chen}
\affiliation{Department of Physics, Yunnan University, Kunming, 650091, China}
\author{Sixia Yu}
\affiliation{Centre for Quantum Technologies, National University of Singapore, 3 Science Drive 2, Singapore 117543, Singapore}
\affiliation{Hefei National Laboratory for Physical Sciences at Microscale and Department of Modern Physics,  University of Science and Technology of China, Hefei, Anhui 230026, China}
\author{Haidong Yuan}
\affiliation{Department of Mechanical and Automation Engineering, The Chinese University of Hong Kong, Hong Kong}
\author{C.H. Oh}
\affiliation{Centre for Quantum Technologies, National University of Singapore, 3 Science Drive 2, Singapore 117543, Singapore}
\affiliation{Physics Department, National University of Singapore, 3 Science Drive 2, Singapore 117543, Singapore}

\begin{abstract}
We propose a hierarchy of Bell-type inequalities for arbitrary $n$-partite systems that identify the different degrees of nonlocality ranging from standard to genuine multipartite nonlocality. After introducing the definition of nonsignaling $m$-locality, we show that the observed joint probabilities in any nonsignaling $m$-local realistic models should satisfy the $(m-1)$-th Bell-type inequality. When $m=2$ the corresponding inequality reduces to the one shown in [Phys. Rev. Lett. 112, 140404 (2014)] whose violation indicates genuine multipartite nonlocality, and when $m=n$ the corresponding inequality is just Hardy's inequality whose violation indicates standard multipartite nonlocality. Furthermore, several examples are provided to demonstrate their hierarchy of multipartite nonlocality.
\end{abstract}
\date{\today}

\maketitle

\section{Introduction}
In 1964, Bell proved that the predictions of quantum theory for some bipartite quantum states are incompatible with those of deterministic local hidden-variable models by the violation of Bell's inequality, and therefore physical theory of local hidden variables cannot reproduce all of the predictions of quantum mechanics \cite{Bell,rmp}. These quantum states which cannot be described by local hidden-variable models are nonlocal. Subsequently, Clause-Horne-Shimony-Holt (CHSH) inequality was introduced for bipartite systems with two different measurement settings for each observer and two possible outcomes for each measurement \cite{CHSH}. Furthermore, the CHSH inequality was generalized to the Collins-Gisin-Linden-Masser-Popescu (CGLMP) inequality with each measurement having more than two possible outcomes \cite{CGLMP}.
Quantum nonlocality is widely used in quantum information tasks \cite{rmp}, such as making the secure quantum communication \cite{cryptography1,cryptography2}, decreasing the communication complexity \cite{complexity}, randomness generation \cite{randomness}, and measurement-based quantum computation \cite{computation}.

The nonlocality issue for bipartite system is simple, it is either nonlocal or local. However, the situation is dramatically changed for multipartite case, since the structure of multipartite nonlocality is far from a simple extension of the bipartite one. For the multipartite case, quantum nonlocality has much richer and more complex structure. Consider an $n$-partite system, there exist $n-1$ kinds of hierarchical multipartite nonlocality. The first one is the standard (or weakest) multipartite nonlocality which is a natural generalization from Bell's bipartite nonlocality. Many Bell-type inequalities have been proposed for the standard multipartite nonlocality, such as the Mermin-Ardehali-Belinskii-Klyshko (MABK) inequalities \cite{MABK}, the Werner-Wolf-Zukowski-Brukner (WWZB) inequality \cite{WWZB}, Hardy's inequality \cite{Hardy,Hardy2,Hardy3}. The last kind of (or strongest) multipartite nonlocality is genuine multipartite nonlocality, which shows that the nonlocality is truly established among all the parties of the system. The detection of genuine multipartite nonlocality has attracted much interest recently. In 1987, Svetlichny first introduced the notion of genuine multipartite nonlocality, and derived a Bell-type inequality for tripartite systems (i.e. Svetlichny inequality) to test the genuine tripartite nonlocality \cite{Svetlichny}. Moreover, Seevinck and Svetlichny and Collins {\it et al.} independently generalized the Svetlichny inequality from tripartite systems to arbitrary $n$-partite systems \cite{multi}. Recently, the Svetlichny inequality has been generalized to arbitrary $n$-partite higher-dimensional systems \cite{higher}. However, Svetlichny's notion of genuine multipartite nonlocality allows correlations capable of signaling among parties \cite{nonsignaling1}, which would be inconsistent with an operational viewpoint \cite{nonsignaling2}. Fortunately, multipartite nonlocality in the nonsignaling scenario was proposed \cite{nonsignaling1,nonsignaling2,nonsignaling3,nonsignaling4,nonsignaling5,nonsignaling6}, since allowing signaling is incongruous with a physical perspective.

As introduced above, the standard and genuine multipartite nonlocality have been studied in many papers. However, Bell-type inequalities for multipartite nonlocality between the standard and genuine one have never been proposed. In this paper, we propose a hierarchy of Bell-type inequalities for arbitrary $n$-partite systems, which can identify the different degrees of nonlocality ranging from standard to genuine multipartite nonlocality. After introducing the definition of nonsignaling $m$-locality, we show that the observed joint probabilities in any nonsignaling $m$-local realistic models should satisfy the $(m-1)$-th Bell-type inequality. When $m=2$ the corresponding inequality reduces to the one shown in Ref. \cite{nonsignaling5} whose violation indicates genuine multipartite nonlocality, and when $m=n$ the corresponding inequality is just Hardy's inequality whose violation indicates standard multipartite nonlocality. Furthermore, several examples are provided to demonstrate the multipartite nonlocality hierarchy.

\section{Nonsignaling $m$-locality} Consider a system composed of $n$ spacelike separated subsystems that are labeled with the index set $I=\{1,2,\cdots,n\}$. The measurement setting and outcome of the $k$-th subsystem ($k\in I$) are denoted by $M_k$ and $r_k$, respectively. $P(r_I|M_I)$ is the joint probability distribution with $r_I=(r_1,\cdots,r_n)$ and $M_I=(M_1,\cdots,M_n)$, when all $n$ parties use the measurement settings $M_1,\cdots,M_n$ and obtain the results $r_1,\cdots,r_n$.

On the one hand, in a standard local hidden variable model, the joint probability distribution $P(r_I|M_I)$  assumes the following form:
\begin{equation}\label{standard}
P(r_I|M_I)=\int q(\lambda)\prod_{k=1}^{n}P_k(r_k|M_k,\lambda)\mathrm{d}\lambda,
\end{equation}
where $\lambda$ is a shared local hidden variable, $q(\lambda)\geq0$ with $\int q(\lambda)\mathrm{d}\lambda=1$, and $P_k(r_k|M_k,\lambda)$ is the probability of the $k$-th observer measuring observable $M_k$ with outcome $r_k$ for a given local hidden variable $\lambda$ distributed according to $q(\lambda)$. Violation of Eq. (\ref{standard}) indicates the standard (or weakest) multipartite nonlocality.

On the other hand, Svetlichny introduced the genuine multipartite nonlocality, which indicates that the joint probability distribution cannot be written as
\begin{equation}\label{genuine}
    P(r_I|M_I)=\sum_{\alpha}\int  q_{\alpha}(\lambda)P_{\alpha}(r_{\alpha}|M_{\alpha},\lambda)P_{\bar{\alpha}}(r_{\bar{\alpha}}|M_{\bar{\alpha}},\lambda)\mathrm{d}\lambda,
\end{equation}
where $\alpha\neq\emptyset$, $\alpha\subset I$, $\bar{\alpha}=I\setminus \alpha$, and $|\alpha|\leq|\bar{\alpha}|$.
Here, the set $I$ has been divided into arbitrary two  nonempty and disjoint subsets  $\alpha$ and $\bar{\alpha}$, and $P_{\alpha}(r_{\alpha}|M_{\alpha},\lambda)$  denotes the joint probability of all observers $k \in\alpha$ measuring observable $M_k$ with outcome $r_k$ for a given local hidden variable $\lambda$ distributed according to $q_{\alpha}(\lambda)$. In the nonsignaling scenario, genuine multipartite nonlocality needs to obey the nonsignaling condition, i.e., $P_{\alpha}(r_{\alpha}|M_{\alpha},\lambda)$ and $P_{\bar{\alpha}}(r_{\bar{\alpha}}|M_{\bar{\alpha}},\lambda)$ in Eq. (\ref{genuine}) satisfying
\begin{eqnarray}\label{nonsignaling}
    \sum_{r_k}P_{\beta}(r_{\beta\setminus k}r_k|M_{\beta\setminus k}M_k,\lambda)&=&\sum_{r_k}P_{\beta}(r_{\beta\setminus k}r_k|M_{\beta\setminus k}M'_k,\lambda)\nonumber\\
    &:=&P_{\beta}(r_{\beta\setminus k}|M_{\beta\setminus k},\lambda)
\end{eqnarray}
for all $k\in\beta$, $\beta=\alpha$ or $\bar{\alpha}$, and $|\beta|\geq2$.

Violations of Eq. (\ref{standard}) and Eq. (\ref{genuine}) indicate the standard (or weakest) and genuine (or strongest) multipartite nonlocality, respectively. Actually, there exist hierarchal kinds of multipartite nonlocality between the standard and the genuine one. We define the $m$-locality in the nonsignaling scenario as follows:

{\bf Definition.} In a nonsignaling $m$-local hidden variable model ($2\leq m\leq n$), the joint probability distribution $P(r_I|M_I)$ assumes the following form:
\begin{equation}\label{m-local}
    P(r_I|M_I)=\sum_{\{\alpha_i\}}\int  q_{\{\alpha_i\}}(\lambda)\prod_{i=1}^{m}P_{\alpha_i}(r_{\alpha_i}|M_{\alpha_i},\lambda)\mathrm{d}\lambda,
\end{equation}
where $\alpha_i\neq\emptyset$, $\alpha_i\subset I$, $\bigcup_{i=1}^{m}\alpha_i=I$, $|\alpha_1|\leq|\alpha_2|\leq\cdots\leq|\alpha_m|$, and $\alpha_i\bigcap\alpha_j=\emptyset$ for $i,j=1,\cdots,m$ and $i\neq j$. If $|\alpha_i|\geq2$, $P_{\alpha_i}(r_{\alpha_i}|M_{\alpha_i},\lambda)$ should satisfy the nonsignaling condition Eq. (\ref{nonsignaling}) for all $k\in\alpha_i$.

Here, the set $I$ has been divided into arbitrary  $m$ nonempty and disjoint subsets  $\alpha_i$ for $i=1,\cdots,m$, and $P_{\alpha_i}(r_{\alpha_i}|M_{\alpha_i},\lambda)$  denotes the joint probability of all observers $k \in\alpha_i$ measuring observable $M_k$ with outcome $r_k$ for a given local hidden variable $\lambda$ distributed according to $q_{\{\alpha_i\}}(\lambda)$. The sum in Eq. (\ref{m-local}) is taken over all possible partitions of $I$ into $m$ nonempty subsets. When $m=2$, Eq. (\ref{m-local}) reduces to Eq. (\ref{genuine}) whose violation indicates genuine multipartite nonlocality. When $m=n$, Eq. (\ref{m-local}) reduces to Eq. (\ref{standard}) whose violation indicates standard  multipartite nonlocality.

\section{Hierarchy of Bell-type inequalities}
As introduced in Sec. I, a Hardy-like Bell-type inequality has been proposed in Ref. \cite{nonsignaling5}:
\begin{eqnarray}
&&P(0_I|a_I)-\sum_{k\in I}P(0_I|b_k a_{\bar{k}})\nonumber\\
&&\ \ \ \ \ \ \  -\sum_{k\in I\setminus\{k'\}}P(1_{k'}1_k 0_{\overline{k'k}}|b_{k'}b_{k}a_{\overline{k'k}})\leq0,\label{chen}
\end{eqnarray}
where the $k$-th local observer measures two alternative observables $\{a_k,b_k\}$ with two outcomes labeled by $\{0,1\}$, $\bar{k}=I\setminus\{k\}$, $\overline{k'k}=I\setminus\{k,k'\}$, and $k'\in I$ is fixed.
All nonsignaling 2-local hidden variable models ($m=2$) satisfy this inequality, and the violation indicates genuine multipartite nonlocality.
Moreover, Hardy's inequality has been proposed for $n$-local hidden variable models ($m=n$) \cite{Hardy,Hardy2,Hardy3}:
\begin{eqnarray}
P(0_I|a_I)-\sum_{k\in I}P(0_I|b_k a_{\bar{k}})-P(1_I|b_I)\leq0,\label{hardy}
\end{eqnarray}
and the violation indicates standard multipartite nonlocality. However, when $2<m<n$ Bell-type inequalities for nonsignaling $m$-local hidden variable models are still missing. In the following, we will present those Bell-type inequalities for nonsignaling $m$-local hidden variable models with $2\leq m\leq n$.

\begin{widetext}
{\bf Theorem.} In any nonsignaling $m$-local hidden variable model ($2\leq m\leq n$), the joint outcome probabilities should satisfy the following $(m-1)$-th Bell-type inequality:
\begin{eqnarray}\label{m-bell}
P(0_I|a_I)-\sum_{k\in I}P(0_I|b_k a_{\bar{k}})
 -\sum\limits_{k_1,\cdots,k_{m-1}\in I\setminus\{k'\} \atop 1\leq k_1<\cdots< k_{m-1}\leq n} P(1_{k' k_1 \cdots k_{m-1}} 0_{\overline{k'k_1\cdots k_{m-1}}}|b_{k' k_1\cdots k_{m-1}}a_{\overline{k'k_1\cdots k_{m-1}}})\leq0,
\end{eqnarray}
where the $k$-th local observer measures two alternative observables $\{a_k,b_k\}$ with two outcomes labeled by $\{0,1\}$, $\bar{k}=I\setminus\{k\}$,
$k' k_1 \cdots k_{m-1}=\{k', k_1, \cdots, k_{m-1}\}$,
$\overline{k'k_1\cdots k_{m-1}}=I\setminus\{k', k_1, \cdots, k_{m-1}\}$, and $k'\in I$ is fixed.

\textbf{Proof.} By linearity of Eq. (\ref{m-local}), we only have to prove that the correlation $\prod_{i=1}^{m}P_{\alpha_i}(r_{\alpha_i}|M_{\alpha_i},\lambda)$ satisfies the inequality Eq.(\ref{m-bell}) for any given $\{\alpha_i\}$ and $\lambda$. It is worth noticing that $k'\in I$ is fixed but can be an arbitrary number with $1\leq k'\leq n$. Since every $\alpha_i$ is a nonempty subset of $I$, without loss of generality, we suppose $k'\in \alpha_1$ and each of the rest $\alpha_i$ ($2\leq i\leq m$) contains at least one element denoted as $j_i$ ($2\leq i\leq m$). The mathematical induction method will be used in the following proof.

(i) When $m=2$, we substitute the probability distribution $P_{\alpha_1}(r_{\alpha_1}|M_{\alpha_1},\lambda)P_{\alpha_2}(r_{\alpha_2}|M_{\alpha_2},\lambda)$ into the left hand side (LHS) of Eq. (\ref{m-bell}). Thus, we have \cite{nonsignaling5}
\begin{eqnarray}
&&\mathrm{LHS}\nonumber\\
&=&P_{\alpha_1}(0_{\alpha_1}|a_{\alpha_1},\lambda)P_{\alpha_2}(0_{\alpha_2}|a_{\alpha_2},\lambda)-\sum_{k\in\alpha_1}P_{\alpha_1}(0_k0_{\alpha_1\setminus k}|b_k a_{\alpha_1\setminus k},\lambda)P_{\alpha_2}(0_{\alpha_2}|a_{\alpha_2},\lambda)\nonumber\\
&&-\sum_{k\in\alpha_2}P_{\alpha_1}(0_{\alpha_1}|a_{\alpha_1},\lambda)P_{\alpha_2}(0_k0_{\alpha_2\setminus k}|b_k a_{\alpha_2\setminus k},\lambda)-\sum_{k\in\alpha_1\setminus k'}P_{\alpha_1}(1_{k'}1_k 0_{\alpha_1\setminus\{k',k\}}|b_{k'}b_k a_{\alpha_1\setminus\{k',k\}},\lambda)P_{\alpha_2}(0_{\alpha_2}|a_{\alpha_2},\lambda)\nonumber\\
&&-\sum_{k\in\alpha_2}P_{\alpha_1}(1_{k'}0_{\alpha_1\setminus k'}|b_{k'}a_{\alpha_1\setminus k'},\lambda)P_{\alpha_2}(1_{k}0_{\alpha_2\setminus k}|b_{k}a_{\alpha_2\setminus k},\lambda)\nonumber\\
&\leq&P_{\alpha_1}(0_{\alpha_1}|a_{\alpha_1},\lambda)P_{\alpha_2}(0_{\alpha_2}|a_{\alpha_2},\lambda)-P_{\alpha_1}(0_{k'}0_{\alpha_1\setminus k'}|b_{k'} a_{\alpha_1\setminus k'},\lambda)P_{\alpha_2}(0_{\alpha_2}|a_{\alpha_2},\lambda)\nonumber\\
&&-P_{\alpha_1}(0_{\alpha_1}|a_{\alpha_1},\lambda)P_{\alpha_2}(0_{j_2}0_{\alpha_2\setminus j_2}|b_{j_2} a_{\alpha_2\setminus j_2},\lambda)-P_{\alpha_1}(1_{k'}0_{\alpha_1\setminus k'}|b_{k'} a_{\alpha_1\setminus k'},\lambda)P_{\alpha_2}(1_{j_2}0_{\alpha_2\setminus j_2}|b_{j_2} a_{\alpha_2\setminus j_2},\lambda)\nonumber\\
&\leq&\bigg(P_{\alpha_1}(0_{\alpha_1}|a_{\alpha_1},\lambda)-P_{\alpha_1}(0_{k'}0_{\alpha_1\setminus k'}|b_{k'} a_{\alpha_1\setminus k'},\lambda)-\min{[P_{\alpha_1}(0_{\alpha_1}|a_{\alpha_1},\lambda),P_{\alpha_1}(1_{k'}0_{\alpha_1\setminus k'}|b_{k'} a_{\alpha_1\setminus k'},\lambda)]}\bigg)P_{\alpha_2}(0_{\alpha_2}|a_{\alpha_2},\lambda)\nonumber\\
&=&\left\{
\begin{array}{ll}
-P_{\alpha_1}(0_{k'}0_{\alpha_1\setminus k'}|b_{k'} a_{\alpha_1\setminus k'},\lambda)P_{\alpha_2}(0_{\alpha_2}|a_{\alpha_2},\lambda)\leq0,   \ \ \ \mathrm{if} \ P_{\alpha_1}(0_{\alpha_1}|a_{\alpha_1},\lambda)\leq P_{\alpha_1}(1_{k'}0_{\alpha_1\setminus k'}|b_{k'} a_{\alpha_1\setminus k'},\lambda);\\
-P_{\alpha_1}(1_{k'}0_{\alpha_1\setminus k'}|a_{k'} a_{\alpha_1\setminus k'},\lambda)P_{\alpha_2}(0_{\alpha_2}|a_{\alpha_2},\lambda)\leq0,   \ \ \ \mathrm{if} \
P_{\alpha_1}(0_{\alpha_1}|a_{\alpha_1},\lambda)>P_{\alpha_1}(1_{k'}0_{\alpha_1\setminus k'}|b_{k'} a_{\alpha_1\setminus k'},\lambda).
\end{array}%
\right.
\end{eqnarray}
The above proof was given in the appendix of Ref. \cite{nonsignaling5}.

(ii) Suppose the inequality holds for $m=t-1$ ($3\leq t\leq n$),
\begin{eqnarray}
&&\mathrm{LHS}\leq\prod_{i=1}^{t-1}P_{\alpha_i}(0_{\alpha_i}|a_{\alpha_i},\lambda)-\sum_{i=2}^{t-1}P_{\alpha_i}(0_{j_i}0_{\alpha_i\setminus j_i}|b_{j_i} a_{\alpha_i\setminus j_i},\lambda)\prod_{i'=1,i'\neq i}^{t-1}P_{\alpha_{i'}}(0_{\alpha_{i'}}|a_{\alpha_{i'}},\lambda)\nonumber\\
&&-P_{\alpha_1}(0_{k'}0_{\alpha_1\setminus k'}|b_{k'} a_{\alpha_1\setminus k'},\lambda)\prod_{i=2}^{t-1}P_{\alpha_i}(0_{\alpha_i}|a_{\alpha_i},\lambda)-P_{\alpha_1}(1_{k'}0_{\alpha_1\setminus k'}|b_{k'} a_{\alpha_1\setminus k'},\lambda)\prod_{i=2}^{t-1}P_{\alpha_i}(1_{j_i}0_{\alpha_i\setminus j_i}|b_{j_i} a_{\alpha_i\setminus j_i},\lambda)\leq0.
\end{eqnarray}
It is worth noticing that when $t=3$ (i.e. $m=2$), this inequality reduces to case (i).

(iii) We now prove the inequality holds for $m=t$  ($3\leq t\leq n$),
\begin{eqnarray}
\mathrm{LHS}&\leq&\prod_{i=1}^{t}P_{\alpha_i}(0_{\alpha_i}|a_{\alpha_i},\lambda)-\sum_{i=2}^{t}P_{\alpha_i}(0_{j_i}0_{\alpha_i\setminus j_i}|b_{j_i} a_{\alpha_i\setminus j_i},\lambda)\prod_{i'=1,i'\neq i}^{t}P_{\alpha_{i'}}(0_{\alpha_{i'}}|a_{\alpha_{i'}},\lambda)\nonumber\\
&&-P_{\alpha_1}(0_{k'}0_{\alpha_1\setminus k'}|b_{k'} a_{\alpha_1\setminus k'},\lambda)\prod_{i=2}^{t}P_{\alpha_i}(0_{\alpha_i}|a_{\alpha_i},\lambda)-P_{\alpha_1}(1_{k'}0_{\alpha_1\setminus k'}|b_{k'} a_{\alpha_1\setminus k'},\lambda)\prod_{i=2}^{t}P_{\alpha_i}(1_{j_i}0_{\alpha_i\setminus j_i}|b_{j_i} a_{\alpha_i\setminus j_i},\lambda)\nonumber\\
&\leq&\bigg(\prod_{i=1}^{t-1}P_{\alpha_i}(0_{\alpha_i}|a_{\alpha_i},\lambda)-\sum_{i=2}^{t-1}P_{\alpha_i}(0_{j_i}0_{\alpha_i\setminus j_i}|b_{j_i} a_{\alpha_i\setminus j_i},\lambda)\prod_{i'=1,i'\neq i}^{t-1}P_{\alpha_{i'}}(0_{\alpha_{i'}}|a_{\alpha_{i'}},\lambda)\nonumber\\
&&-P_{\alpha_1}(0_{k'}0_{\alpha_1\setminus k'}|b_{k'} a_{\alpha_1\setminus k'},\lambda)\prod_{i=2}^{t-1}P_{\alpha_i}(0_{\alpha_i}|a_{\alpha_i},\lambda)\nonumber\\
&&-\min{\Big[\prod_{i=1}^{t-1}P_{\alpha_i}(0_{\alpha_i}|a_{\alpha_i},\lambda),P_{\alpha_1}(1_{k'}0_{\alpha_1\setminus k'}|b_{k'} a_{\alpha_1\setminus k'},\lambda)\prod_{i=2}^{t-1}P_{\alpha_i}(1_{j_i}0_{\alpha_i\setminus j_i}|b_{j_i} a_{\alpha_i\setminus j_i},\lambda)\Big]}\bigg)P_{\alpha_t}(0_{\alpha_t}|a_{\alpha_t},\lambda).
\end{eqnarray}
If $\prod_{i=1}^{t-1}P_{\alpha_i}(0_{\alpha_i}|a_{\alpha_i},\lambda)\leq P_{\alpha_1}(1_{k'}0_{\alpha_1\setminus k'}|b_{k'} a_{\alpha_1\setminus k'},\lambda)\prod_{i=2}^{t-1}P_{\alpha_i}(1_{j_i}0_{\alpha_i\setminus j_i}|b_{j_i} a_{\alpha_i\setminus j_i},\lambda)$, then we have $\mathrm{LHS}\leq -\big(\sum_{i=2}^{t-1}P_{\alpha_i}(0_{j_i}0_{\alpha_i\setminus j_i}|b_{j_i} a_{\alpha_i\setminus j_i},\lambda)\prod_{i'=1,i'\neq i}^{t-1}P_{\alpha_{i'}}(0_{\alpha_{i'}}|a_{\alpha_{i'}},\lambda)+P_{\alpha_1}(0_{k'}0_{\alpha_1\setminus k'}|b_{k'} a_{\alpha_1\setminus k'},\lambda)\prod_{i=2}^{t-1}P_{\alpha_i}(0_{\alpha_i}|a_{\alpha_i},\lambda)\big)\times P_{\alpha_t}(0_{\alpha_t}|a_{\alpha_t},\lambda)\leq0$. If $\prod_{i=1}^{t-1}P_{\alpha_i}(0_{\alpha_i}|a_{\alpha_i},\lambda)> P_{\alpha_1}(1_{k'}0_{\alpha_1\setminus k'}|b_{k'} a_{\alpha_1\setminus k'},\lambda)\prod_{i=2}^{t-1}P_{\alpha_i}(1_{j_i}0_{\alpha_i\setminus j_i}|b_{j_i} a_{\alpha_i\setminus j_i},\lambda)$, then LHS reduces to case (ii) ($m=t-1$) with a nonnegative factor $P_{\alpha_t}(0_{\alpha_t}|a_{\alpha_t},\lambda)$, thus $\mathrm{LHS}\leq0$.

Therefore, Eq. (\ref{m-bell}) holds in any nonsignaling $m$-local hidden variable model ($2\leq m\leq n$).     \hfill  $\blacksquare$

\textbf{Remark.}  When $m=2$ and $m=n$, Eq. (\ref{m-bell}) reduces to Eq. (\ref{chen}) and Eq. (\ref{hardy}), respectively. In Eq. (\ref{m-bell}), there are $n$ terms in the first sum and ${\binom{n-1}{m-1}}$ terms in the second sum. All nonsignaling $m$-local models will satisfy Eq. (\ref{m-bell}). For an arbitrary $n$-particle quantum state $\varrho$, to violate Eq. (\ref{m-bell}) one must find two measurement settings $\{|a_i\rangle,|b_i\rangle\}$ for each particle $i$,
\begin{eqnarray}
&&\langle a_I|\varrho|a_I\rangle-\sum_{k\in I}\langle b_k a_{\bar{k}}|\varrho|b_k a_{\bar{k}}\rangle
-\sum_{2\leq k_1<\cdots< k_{m-1}\leq n}\langle \bar{b}_1\bar{b}_{k_1}\cdots\bar{b}_{k_{m-1}}a_{\overline{1k_1\cdots k_{m-1}}}|\varrho| \bar{b}_1\bar{b}_{k_1}\cdots\bar{b}_{k_{m-1}}a_{\overline{1k_1\cdots k_{m-1}}}\rangle>0,
\end{eqnarray}
where $|a_{\alpha}\rangle=\otimes_{k\in\alpha}|a_k\rangle$ and $|\bar{b}_k\rangle$ is orthogonal to $|b_k\rangle$. For simplicity we just set $k'=1$.

\end{widetext}

\section{Examples}
In this section, we will present several examples to demonstrate their multipartite nonlocality hierarchy. For example, when $n=4$, there are 3 Bell-type inequalities in Eq. (\ref{m-bell}). To violate them, one must find two measurement settings $\{|a_i\rangle,|b_i\rangle\}$ for each particle $i$ such that,
\begin{eqnarray}
&&\langle \hat{a}_1 \hat{a}_2 \hat{a}_3 \hat{a}_4\rangle-\langle \hat{b}_1 \hat{a}_2 \hat{a}_3 \hat{a}_4\rangle-\langle \hat{a}_1 \hat{b}_2 \hat{a}_3 \hat{a}_4\rangle\nonumber\\
&&-\langle \hat{a}_1 \hat{a}_2 \hat{b}_3 \hat{a}_4\rangle
-\langle \hat{a}_1 \hat{a}_2 \hat{a}_3 \hat{b}_4\rangle-\langle \hat{\bar{b}}_1 \hat{\bar{b}}_2 \hat{a}_3 \hat{a}_4\rangle\nonumber\\
&&-\langle \hat{\bar{b}}_1 \hat{a}_2 \hat{\bar{b}}_3 \hat{a}_4\rangle-\langle \hat{\bar{b}}_1 \hat{a}_2 \hat{a}_3 \hat{\bar{b}}_4\rangle>0,\label{4-genuine}\\
&&\langle \hat{a}_1 \hat{a}_2 \hat{a}_3 \hat{a}_4\rangle-\langle \hat{b}_1 \hat{a}_2 \hat{a}_3 \hat{a}_4\rangle-\langle \hat{a}_1 \hat{b}_2 \hat{a}_3 \hat{a}_4\rangle\nonumber\\
&&-\langle \hat{a}_1 \hat{a}_2 \hat{b}_3 \hat{a}_4\rangle
-\langle \hat{a}_1 \hat{a}_2 \hat{a}_3 \hat{b}_4\rangle-\langle \hat{\bar{b}}_1 \hat{\bar{b}}_2 \hat{\bar{b}}_3 \hat{a}_4\rangle\nonumber\\
&&-\langle \hat{\bar{b}}_1 \hat{a}_2 \hat{\bar{b}}_3 \hat{\bar{b}}_4\rangle-\langle \hat{\bar{b}}_1 \hat{\bar{b}}_2 \hat{a}_3 \hat{\bar{b}}_4\rangle>0,\label{4-h}\\
&&\langle \hat{a}_1 \hat{a}_2 \hat{a}_3 \hat{a}_4\rangle-\langle \hat{b}_1 \hat{a}_2 \hat{a}_3 \hat{a}_4\rangle-\langle \hat{a}_1 \hat{b}_2 \hat{a}_3 \hat{a}_4\rangle\nonumber\\
&&-\langle \hat{a}_1 \hat{a}_2 \hat{b}_3 \hat{a}_4\rangle
-\langle \hat{a}_1 \hat{a}_2 \hat{a}_3 \hat{b}_4\rangle-\langle \hat{\bar{b}}_1 \hat{\bar{b}}_2 \hat{\bar{b}}_3 \hat{\bar{b}}_4\rangle>0,\label{4-s}
\end{eqnarray}
where $\hat{a}_k=|a_k\rangle\langle a_k|$, $\hat{b}_k=|b_k\rangle\langle b_k|$, $\hat{\bar{b}}_k=\idol-|b_k\rangle\langle b_k|$. Eqs. (\ref{4-genuine}) and (\ref{4-s}) indicate the genuine and standard multipartite nonlocality, respectively. Our new Bell-type inequality is Eq. (\ref{4-h}) which indicates the nonlocality between the genuine and the standard one.

\begin{table}[tb]
\renewcommand\arraystretch{1.3}
\centering
\caption{\label{Table:1} Numerical search results of $p_i$ for $\varrho_{\mathrm{GHZ}}=p|\mathrm{GHZ}\rangle\langle \mathrm{GHZ}|+(1-p)\frac{\idol_n}{2^n}$ with $n=4,5,6$.  When $p>p_i$ ($1\leq i\leq n-1$), the noisy GHZ state $\varrho_{\mathrm{GHZ}}$ will violate the $i$-th Bell-type inequality in Eq. (\ref{m-bell}) derived from nonsignaling $(i+1)$-local hidden variable models.}

\begin{tabular}{c c c c c c} \hline
\ \ \ $n$\ \ \ &\ \ \ $p_1$\ \ \ &\ \ \  $p_2$\ \ \  &\ \ \ $p_3$ \ \ \ & \ \ \ $p_4$ \ \ \ & \ \ \ $p_5$ \ \ \   \\\hline
4            &$0.948$     &$0.914$    &0.822         &     -    & -   \\
5            &$0.964$     &$0.952$    &0.923         &0.847         & -   \\
6            &$0.971$      &$0.969$    &0.960         &0.931         &0.866    \\\hline
\end{tabular}
\end{table}

We consider $n$-qubit mixed states, i.e., the noisy GHZ state and W state,
\begin{eqnarray}
\varrho_{\mathrm{GHZ}}&=&p|\mathrm{GHZ}\rangle\langle \mathrm{GHZ}|+(1-p)\frac{\idol_n}{2^n},\\
\varrho_{\mathrm{W}}&=&p|W\rangle\langle W|+(1-p)\frac{\idol_n}{2^n},
\end{eqnarray}
where $\idol_n$ is a $2^n\times 2^n$ identity matrix, $|\mathrm{GHZ}\rangle=(|00\cdots0\rangle+|11\cdots1\rangle)/\sqrt{2}$, $|W\rangle=(\sum_{\mathrm{perm}}|0\cdots01\rangle)/\sqrt{n}$ with the sum taking over all possible permutation cases of 1 one and $n-1$ zeros. For simplicity, we choose all the measurement settings in the X-Z plane of the Bloch Sphere and assume $\hat{a}_k=\hat{a}$, $\hat{b}_k=\hat{b}$ for all $2\leq k\leq n$ due to the symmetry of the inequalities. After numerical search for $n=4,5,6$, we find that when $p>p_i$ ($1\leq i\leq n-1$) the corresponding state $\varrho_{\mathrm{GHZ}}$ or $\varrho_{\mathrm{W}}$ will violate the $i$-th Bell-type inequality in Eq. (\ref{m-bell}) derived from nonsignaling $(i+1)$-local hidden variable models, which demonstrates hierarchical multipartite nonlocality for different $p_i$. All the $p_i$ are listed in Table \ref{Table:1} and Table \ref{Table:2} for $\varrho_{\mathrm{GHZ}}$ and $\varrho_{\mathrm{W}}$ with $n=4,5,6$, respectively.

\begin{table}[tb]
\renewcommand\arraystretch{1.3}
\centering
\caption{\label{Table:2} Numerical search results of $p_i$ for $\varrho_{\mathrm{W}}=p|W\rangle\langle W|+(1-p)\frac{\idol_n}{2^n}$ with $n=4,5,6$.  When $p>p_i$ ($1\leq i\leq n-1$),the noisy W state $\varrho_{\mathrm{W}}$ will violate the $i$-th Bell-type inequality in Eq. (\ref{m-bell}) derived from nonsignaling $(i+1)$-local hidden variable models.}

\begin{tabular}{c c c c c c} \hline
\ \ \ $n$\ \ \ &\ \ \ $p_1$\ \ \ &\ \ \  $p_2$\ \ \  &\ \ \ $p_3$ \ \ \ & \ \ \ $p_4$ \ \ \ & \ \ \ $p_5$ \ \ \   \\\hline
4            &$0.903$     &$0.770$    & 0.573        &     -    & -   \\
5            &$0.911$       &$0.792$    & 0.688        &  0.462       & -   \\
6            &$0.894$      &$0.783$    & 0.721        &  0.593       &0.344    \\\hline
\end{tabular}
\end{table}

\section{Discussions and conclusion}
It is worth noticing that the nonsigaling $m$-local hidden variable models are related with $k$-separable states. An $n$-partite pure quantum state $|\Psi_{k-sep}\rangle$ is called $k$-separable \cite{k-sep1,k-sep2,k-sep3,k-sep4,k-sep5,k-sep6,k-sep7,k-sep8,k-sep9,k-sep10,k-sep11,k-sep12,k-sep13,k-sep14,k-sep15,k-sep16,k-sep17,k-sep18,k-sep19,k-sep20}, if and only if there is a $k$-partition $\alpha_1|\alpha_2|\cdots|\alpha_k$ such that  $|\Psi_{k-sep}\rangle$ can be written as a product of $k$ substates:
\begin{eqnarray}
|\Psi_{k-sep}\rangle=|\psi_1\rangle_{\alpha_1}\otimes|\psi_2\rangle_{\alpha_2}\otimes\cdots\otimes|\psi_k\rangle_{\alpha_k},
\end{eqnarray}
where the set $I=\{1,2,\cdots,n\}$ has been split into arbitrary $k$ nonempty and disjoint subsets  $\alpha_i$, i.e.,  $\alpha_i\neq\emptyset$, $\alpha_i\subset I$, $\bigcup_{i=1}^{k}\alpha_i=I$, and $\alpha_i\bigcap\alpha_j=\emptyset$ for $i\neq j$. A $n$-partite mixed state $\varrho_{k-sep}$ is called $k$-separable, if and only if it can be written as a convex combination of $k$-separable pure states:
\begin{eqnarray}
\varrho_{k-sep}=\sum_i p_i|\Psi_{k-sep}^i\rangle\langle\Psi_{k-sep}^i|,\label{k-sep}
\end{eqnarray}
where $|\Psi_{k-sep}^i\rangle$ might be $k$-separable under different $k$-partitions. Comparing Eq. (\ref{m-local}) with Eq. (\ref{k-sep}), one can conclude that all $k$-separable states are nonsignaling $k$-local, since we can always find a nonsignaling $k$-local hidden variable model to describe the joint probability distribution from any $k$-separable states. This is similar to that all bipartite separable states are bipartite local. Therefore, violation of Eq. (\ref{m-bell}) guarantees its $m$-nonseparable, i.e., it is a sufficient condition for detecting $m$-nonseparable states.

In conclusion, we have proposed a hierarchy of Bell-type inequalities for arbitrary $n$-partite systems, which can identify the different degrees of nonlocality ranging from standard to genuine multipartite nonlocality. After introducing the definition of nonsignaling $m$-locality, we have shown that the observed joint probabilities in any nonsignaling $m$-local realistic models should satisfy the $(m-1)$-th Bell-type inequality. When $m=2$ the corresponding inequality reduces to the one shown in Ref. \cite{nonsignaling5} whose violation indicates genuine multipartite nonlocality, and when $m=n$ the corresponding inequality is just Hardy's inequality whose violation indicates standard multipartite nonlocality. Furthermore, several examples have been provided to demonstrate the multipartite nonlocality hierarchy, and the relations between $m$-locality and $k$-separable states have been discussed.

\section*{ACKNOWLEDGMENTS}
This work is funded by the Singapore Ministry of Education (partly through the Academic Research Fund Tier 3 MOE2012-T3-1-009), the National Research Foundation, Singapore (Grant No. WBS: R-710-000-008-271), the financial support from RGC of Hong Kong(Grant No. 538213), the National Natural Science Foundation of China (Grants No. 11504253 and No. 11575155), the open funding program from Key Laboratory of Quantum Information, CAS (Grant No. KQI201605), and the startup funding from Soochow University (Grant No. Q410800215).


\end{document}